\documentclass{article}

\usepackage{amsmath, amssymb, graphicx}

\newcommand{\thetanose}{\theta_\mathrm{nose}}

\newcommand{\half}{1/2}

\def\Xint#1{\mathchoice
{\XXint\displaystyle\textstyle{#1}}%
{\XXint\textstyle\scriptstyle{#1}}%
{\XXint\scriptstyle\scriptscriptstyle{#1}}%
{\XXint\scriptscriptstyle\scriptscriptstyle{#1}}%
\!\int}
\def\XXint#1#2#3{{\setbox0=\hbox{$#1{#2#3}{\int}$}
\vcenter{\hbox{$#2#3$}}\kern-.5\wd0}}

\def\dashint{\Xint-}

\title{Corner and finger formation in Hele--Shaw flow with kinetic undercooling regularisation}

\author{Michael C. Dallaston\footnote{
Mathematical Sciences, Queensland University of Technology, Brisbane 4000, AUSTRALIA
} \ and Scott W. McCue$^*$}

\date{\today}

\begin{document}

\maketitle

\begin{abstract}

We examine the effect of a kinetic undercooling condition on the evolution of a free boundary in Hele--Shaw flow, in both bubble and channel geometries.  We present analytical and numerical evidence that the bubble boundary is unstable and may develop one or more corners in finite time, for both expansion and contraction cases.  This loss of regularity is interesting because it occurs regardless of whether the less viscous fluid is displacing the more viscous fluid, or vice versa.   We show that small contracting bubbles are described to leading order by a well-studied geometric flow rule.  Exact solutions to this asymptotic problem continue past the corner formation until the bubble contracts to a point as a slit in the limit.   Lastly, we consider the evolving boundary with kinetic undercooling in a Saffman--Taylor channel geometry.  The boundary may either form corners in finite time, or evolve to a single long finger travelling at constant speed, depending on the strength of kinetic undercooling.  We demonstrate these two different behaviours numerically.  For the travelling finger, we present results of a numerical solution method similar to that used to demonstrate the selection of discrete fingers by surface tension.  With kinetic undercooling, a continuum of corner-free travelling fingers exists for any finger width above a critical value, which goes to zero as the kinetic undercooling vanishes.
We have not been able to compute the discrete family of analytic solutions, predicted by previous asymptotic analysis, because the numerical scheme cannot distinguish between solutions characterised by analytic fingers and those which are corner-free but non-analytic.
\end{abstract}


\section{Introduction}

The mathematical model of an evolving fluid region in a Hele--Shaw cell is a canonical example of a free boundary problem in applied mathematics.  The amount written on this problem in the last half-century is immeasurable, ranging from experimental work and numerical computations, to explicit solution methods and theoretical results (see the web site~\cite{Howison_website} for an extensive bibliography).

\begin{figure}
\centering
\includegraphics{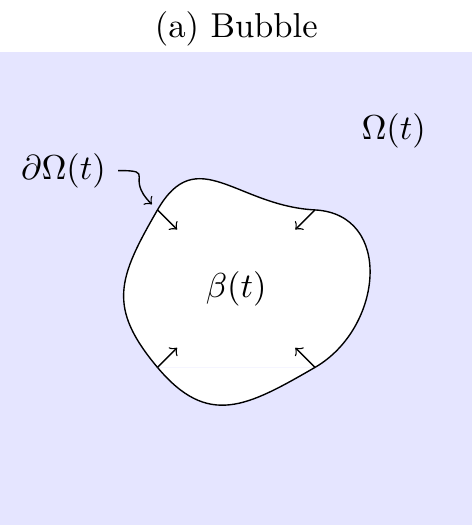}
\qquad
\includegraphics{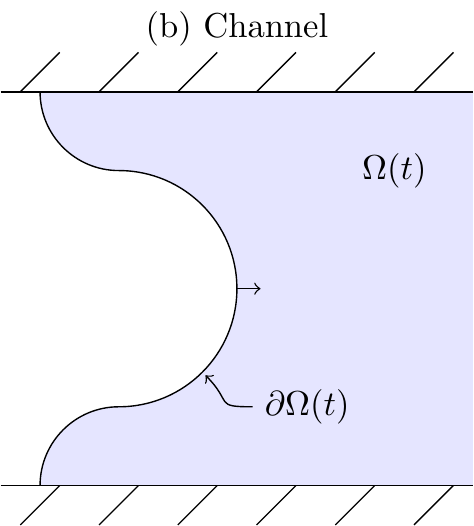}
\caption{A schematic of the evolving fluid region $\Omega(t)$ in free-boundary Hele--Shaw flow, in (a) a bubble geometry, and (b) a channel geometry.}
\label{fig:schematic}
\end{figure}

In this paper we consider a \emph{kinetic undercooling} condition on the boundary $\partial\Omega$ of a contracting or expanding inviscid bubble $\beta(t)$ surrounded by a region of viscous fluid $\Omega(t) \subset \mathbb R^2$, {as well as the equivalent problem in a channel geometry} (see Figure~\ref{fig:schematic}).  The kinetic undercooling condition states that the potential at a point on the free boundary is proportional to the normal velocity of the boundary at that point. While the name derives from the physics of melting and freezing (Stefan) problems~\cite{Evans2000, King2005, Back2014}, similar boundary conditions arise from modelling changes in curvature in the neglected transverse dimension in the fluid dynamics context~\cite{Romero_phd}, diffusion of solvent into glassy polymers~\cite{Cohen1988, Fasano1986, Mccue2011a, Mitchell2012} and in streamers involved in the formation of sparks and lightning~\cite{Ebert2011, Ebert2007, Luque2008, Meulenbroeck2005, Tanveer2009,Kao2010}.  Mathematically, kinetic undercooling is considered as an alternative to surface tension in \emph{regularising} unstable Hele--Shaw flows, by suppressing the blow-up in the speed of the boundary that characterises the unregularised (constant pressure condition) problem, which is ill-posed~\cite{Howison1992}.  
Numerical and existence results for Hele--Shaw flow with kinetic undercooling have been derived for the problem of a viscous blob surrounded by inviscid fluid~\cite{Pleshchinskii2002,Reissig1999}, a situation we do not consider in the present work.

We have recently considered the interplay between surface tension and kinetic undercooling for a contracting inviscid bubble near extinction~\cite{Dallaston2013_NON}, but are aware of no other published research on expanding or contracting bubbles with kinetic undercooling only.  On the other hand, much has been done on inviscid bubbles in the unregularised case, without surface tension or kinetic undercooling.  Many exact solutions are known~\cite{Crowdy2002, Dallaston2012, Howison1986a, Howison1986b, Howison1992}, and the behaviour of bubbles near extinction and break-up has been studied~\cite{Entov1991, Entov2011, King2009, Lee2006, Mccue2003}.  With surface tension regularisation, there are many sophisticated numerical schemes that model an expanding (unstable) boundary and capture the long fingering patterns that result~\cite{Hou1997,Hou1994}.

  There has also been some research on the problem in a channel geometry; the selection problem of determining the discrete set of allowed widths of a travelling finger for varying kinetic undercooling strength (the analogue of the famous selection problem with surface tension~\cite{Chapman1999, Kessler1988, Mclean1981, Tanveer1987, Vandenbroeck1983}) has been analysed using exponential asymptotics~\cite{Chapman2003}, and some numerical studies have been attempted~\cite{Dallaston2011, Romero_phd}.  The conclusions vary: asymptotics suggests that kinetic undercooling selects discrete fingers whose width tends to $\half$ as kinetic undercooling vanishes, while numerical results suggest either no selection (a continuous family of solutions)~\cite{Romero_phd} or discrete branches whose width vanishes~\cite{Dallaston2011}.  We aim to explain these discrepancies in the current paper.
There are also linear stability results~\cite{Chapman2003,Howison1992} and exact travelling waves for sufficiently large kinetic undercooling~\cite{Chapman2003}.
Closely related to the finger is the problem of a travelling {finite} bubble in a directed flow field, which is of recent interest in the literature on streamers~\cite{Gunther2009, Kao2010, Tanveer2009}.  Numerical solutions of the evolving shape of a bubble suggest that an initially smooth bubble boundary may not remain so for all time~\cite{Kao2010}; this property also arises in the present work.

The aim of this paper is to demonstrate the very different effect that kinetic undercooling has on free boundary Hele--Shaw flow, compared with surface tension and unregularised variants, in the context of expanding or contracting bubbles, and translating fronts or fingers in a channel. Table~\ref{tab:table} summarises these differences with particular regard to expanding and contracting bubbles.  

The content of this paper is as follows.  In Section~\ref{sec:formulation} we outline the governing systems of equations. In Section~\ref{sec:bubble} we examine expanding and contracting inviscid bubbles. With linear stability analysis (\S\ref{sec:stability}) we show that in both cases, kinetic undercooling destabilises a near-circular bubble once its radius is outside a critical interval. We also present strong numerical evidence (\S\ref{sec:numerics}) that this instability generically leads to the formation of \emph{corners} on the boundary.  This corner formation represents a blow-up in the curvature, but not speed, of the boundary.  In contrast, surface tension suppresses singularities in both curvature and speed (corners may exist in the unregularised problem, but they are not generic, and exist at a stationary point for a finite time before smoothing off~\cite{King1995}).

In Section~\ref{sec:smallbubble} we show that the boundary in the limit of small bubble size (equivalently, large kinetic undercooling) behaves to leading order as a plane curve that evolves so that each point on the curve moves with constant normal velocity, once a time scaling is applied.  For contracting bubbles this limit is particularly relevant; the extinction behaviour of a bubble in the full Hele--Shaw problem is given by the extinction behaviour of the boundary contracting according to the simpler constant velocity rule.  We derive the exact solution to the leading order problem for an initially elliptical boundary, which demonstrates the formation of corners and the asymptotic shape of the boundary: a slit of vanishing aspect ratio.  Such behaviour is likely to be generic for a wide class of initial conditions.

In Section~\ref{sec:channel} we consider Hele--Shaw flow with kinetic undercooling in a channel.  Here the pertinent question is the existence of travelling wave solutions: either a front that spans the width of the channel (and usually exhibiting corners), or a finger of width strictly less than the channel.  Explicit solutions have been found for the travelling fronts~\cite{Chapman2003}.  For fingers, we present a numerical scheme, previously described in our conference paper~\cite{Dallaston2011}, which shows that corner-free fingers exist for a continuous range of widths bounded below by a minimum width that vanishes as kinetic undercooling goes to zero. 
While this continuum of solutions appears to contradict the findings of Chapman and King~\cite{Chapman2003}, who show there exist discrete families of analytic fingers, the results are entirely consistent because
the numerical scheme cannot distinguish between solutions characterised by analytic fingers and travelling fingers that are corner-free but not analytic (due to the nonexistence of higher derivatives).  Indeed, Chapman and King discuss the existence of these non-analytic travelling finger solutions, and predict the resulting difficulty of numerically computing the discrete spectrum of analytic solutions.
%
%
These results also explain the conclusion of a previous numerical study~\cite{Romero_phd} that kinetic undercooling offered no discrete selection (the discrete branches we found previously~\cite{Dallaston2011} are the result of small numerical errors).  
In Section~\ref{sec:numnumchannel} we alter our numerical scheme from Section~\ref{sec:numerics} to track the evolution of a boundary to either a front or finger, depending on the kinetic undercooling strength.
We conclude with suggestions for further study in  Section~\ref{sec:discussion}.

\begin{table}
\begin{tabular}{p{0.16\textwidth}|p{0.36\textwidth}p{0.36\textwidth}} \hline
\textsc{Boundary} & \textsc{Expanding} & \textsc{Contracting} \\ \hline
Unregularised & \emph{Ill-posed}: generic cusp formation, corners possible~\cite{Howison1986,Howison1986a,Howison1992,King1995}. & \emph{Stable}: 2nd mode neutrally stable, generic extinction shape an ellipse~\cite{Entov1991} \\
& & \\
Surface tension &  \emph{Unstable} but not ill-posed: high modes stable, fingering solutions exist for all time~\cite{Ceniceros1998,Dallaston2013_NON,Hou1997,Paterson1981}. & \emph{Stable}: All modes stable: generic extinction shape a circle~\cite{Dallaston2013_NON}. \\
& & \\
Kinetic undercooling & \emph{Unstable}: all modes unstable, generic corner formation, only weak solutions subsequently (\S 3)& \emph{Unstable}: corners form, generic extinction shape a slit (\S\S 3,4).\\
\hline 
\end{tabular}
\caption{Numerically and analytically observed properties of contracting and expanding inviscid bubbles in Hele--Shaw flow, with various regularisations. The last row contains the results of this paper.  We consider the competition between surface tension and kinetic undercooling for contracting bubbles elsewhere~\protect{\cite{Dallaston2013_NON}}.}
\label{tab:table}
\end{table}

\section{Model equations}
\label{sec:formulation}
The bubble problem consists of solving both a two-dimensional harmonic velocity potential $\phi(\boldsymbol x,t)$ (where $\boldsymbol x=(x,y)$) in $\Omega(t)$ and evolution of the free boundary $\partial\Omega(t)$ (see Figure~\ref{fig:schematic}).  Two boundary conditions are needed on the free boundary $\partial\Omega$: a kinematic condition that relates the fluid velocity to that of the boundary, and a dynamic condition that, in this case, includes the kinetic undercooling term.  The model has two parameters: the kinetic undercooling  coefficient $c$ (a length scale) and the far-field source strength.  For this geometry we choose to scale time such that the far-field source strength is unity, so that the rate of change of area $\mathcal A$ of the bubble $\beta(t) = \mathbb R^2 \backslash (\Omega\cup\partial\Omega)$ is
\begin{equation}
\frac{\mathrm d\mathcal A}{\mathrm dt} = -2\pi,
\label{eq:area}
\end{equation}
and scale length such that $c=1$.  The system in nondimensional variables is
\begin{subequations}
\label{eq:system}
\begin{align}
&\nabla^2\phi = 0, & \boldsymbol x \in \Omega(t), \label{eq:laplace} \\
&\frac{\partial \phi}{\partial n} = v_n, &  \boldsymbol x \in \partial\Omega(t) \label{eq:k} \\
&\phi = v_n, &  \boldsymbol x \in \partial\Omega(t), \label{eq:d} \\
&\phi \sim -\log|\boldsymbol x|, &|\boldsymbol x|\rightarrow\infty. \label{eq:farfield}
\end{align}
\end{subequations}
The source term (\ref{eq:farfield}) and negative change in area (\ref{eq:area}) are for a bubble contracting as time $t$ increases. However, the Hele--Shaw problem with kinetic undercooling is time-reversible, so expansion may simply be considered by decreasing $t$ instead.

In the channel geometry, the fluid region $\Omega$ is confined to the right of the boundary $\partial\Omega$ and between the channel walls (see Figure~\ref{fig:schematic}).  In this geometry there is an additional length scale: the channel width.  For this geometry we scale lengths such that the channel has width $2$ (say, $-1<y<1$), and reintroduce the kinetic undercooling coefficient $c$.  In addition, we have no-flux conditions on the wall and a constant velocity (scaled to unity) in the far field $x\rightarrow\infty$.  The system of equations is
\begin{subequations}
\label{eq:channelsystem}
\begin{align}
&\nabla^2\phi = 0, & \boldsymbol x \in \Omega(t)  \label{eq:channellaplace}\\
&\frac{\partial \phi}{\partial n} = v_n, &  \boldsymbol x \in \partial\Omega(t) \label{eq:channelk} \\
&\phi = cv_n, &  \boldsymbol x \in \partial\Omega(t), \label{eq:channeld} \\
&\phi_y = 0, & x\in\mathbb R, \ y = \pm 1, \label{eq:channelwalls} \\
&\phi \sim x, &x\rightarrow\infty, \ -1<y<1. \label{eq:channelfarfield}
\end{align}
\end{subequations}
Both problems (\ref{eq:system}) and (\ref{eq:channelsystem}) require an initial condition $\Omega(0)$.  For the nondimensionalised bubble problem (\ref{eq:system}), an initial condition for which the length of the boundary $\partial\Omega(0)$ is $\mathcal O(1)$ corresponds to a bubble whose size is the same order as kinetic undercooling.  It is this scale that we primarily consider in the next section.

\section{Corner formation for expanding and contracting bubbles}
\label{sec:bubble}
In this section we demonstrate that an inviscid bubble evolving according to Hele--Shaw flow with kinetic undercooling (\ref{eq:system}) is unstable, both for expansion and contraction.  Linear stability suggests this property, while numerical results establish that corners may form in both cases.  

\subsection{Linear stability}
\label{sec:stability}
\begin{figure}
\centering
\includegraphics{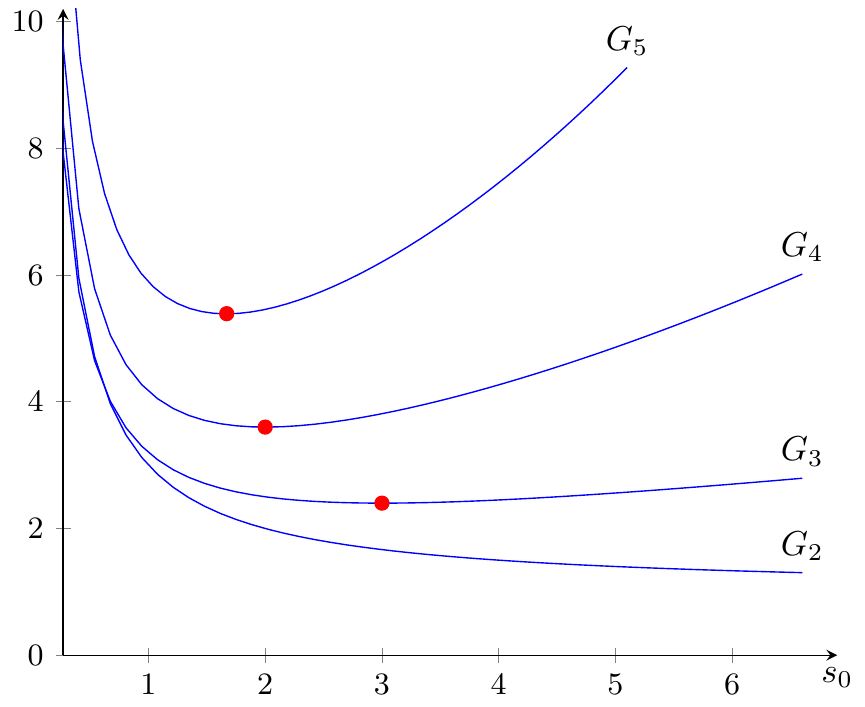}
\caption{A plot of the first few modes of perturbation $G_n, n=2,3,\ldots$, given by (\ref{eq:G}), against the leading order bubble radius $s_0$.  The minimum of each mode is marked (circles).  All modes are unstable for $s_0<1$, and all except the second are unstable for $s_0>3$.  A circular bubble is ultimately unstable for both expansion and contraction, unlike the unregularised or surface tension cases, in which the contracting bubble is stable.}
\label{fig:stability}
\end{figure}

The system (\ref{eq:system}) has an exact, radially symmetric solution where $\partial\Omega$ is a circle of time-dependent radius $s_0(t)$.  In polar coordinates ($\boldsymbol x = r\boldsymbol e_r + \theta\boldsymbol e_\theta$) this exact solution is
\begin{equation}
\phi_0(r,t) = - \log \left(\frac{r}{s_0}\right) +\frac{1}{s_0}, \qquad s_0(t) = \sqrt{s_0(0)^2 - 2t},
\end{equation}
where $r=s_0(t)$ on the free boundary $\partial\Omega$.  We have previously considered the stability of this exact solution to infinitesimal perturbations~\cite{Dallaston2013_NON}; similar approaches have been widely applied to bubbles in Hele--Shaw flow with various boundary conditions and fluid properties~\cite{Paterson1981,Mccue2011,Martyushev2008,Rocha2013,Dias2010,Dias2012}.  Since the circle radius $s_0$ is monotonic in time $t$ we can treat it as the time-like variable.  We perturb the potential $\phi$ and bubble boundary $r = s(\theta,t)$ by an $n$th mode term:
\begin{equation}
\phi \sim \phi_0 + \epsilon A_n(s_0)r^{-n}\cos n\theta, \qquad s\sim s_0 + \epsilon \gamma_n(s_0)\cos n\theta, \qquad \epsilon \ll 1,
\end{equation}
where we must solve for $\gamma_n$ and $A_n$.  The stability of the evolving circle is determined by the change in relative magnitude $G_n(s_0) = \gamma_n(s_0)/s_0$.  By taking the $\mathcal O(\epsilon)$ terms in (\ref{eq:system}) and solving an ordinary differential equation for $\gamma_n$, we find that $G_n$ is given by
\begin{equation}
G_n(s_0) = G_n(1)\frac{(s_0 + n)^{n-1}}{s_0(1+n)^{n-1}}.
\label{eq:G}
\end{equation}
We plot the evolution of the modes over $s_0$ in Figure~\ref{fig:stability}.  

If $G_n$ is increasing in $s_0$ then the $n$th mode is unstable for expanding bubbles and stable for contracting bubbles, while if $G_n$ is decreasing in $s_0$, the opposite holds.  
We disregard the first mode $n=1$ as it corresponds to a translation, which is neutrally stable in absolute terms ($\gamma_1 = $  constant).  
The second mode $G_2$ increases without bound as $s_0\rightarrow 0$ and vanishes as $s_0\rightarrow\infty$.  For all other modes, $G_n$ has an absolute minimum at $s_0= n/(n-2)$, and increases without bound as $s_0$ tends to zero and infinity.  The circle is ultimately unstable for expanding and contracting bubbles, with all modes unstable for $s_0<1$ and all except the second mode $G_2$ unstable for $s_0>3$.  

While all modes $n\geq 3$ are unstable for both $s_0\rightarrow 0$ and $s_0 \rightarrow\infty$, the stability analysis does indicate that kinetic undercooling removes the ill-posedness of unregularised Hele--Shaw flow, since for a fixed $s_0$ the growth of modes is bounded as $n\rightarrow\infty$, that is
\[
G_n(s_0)/G_n(1) \sim 1/s_0, \qquad n\rightarrow\infty.
\]
This moderation of the growth of high order modes also occurs in the stability analysis of Hele--Shaw flow with kinetic undercooling in a channel~\cite{Chapman2003}.  In contrast, in the unregularised (ill-posed) problem the growth of modes is unbounded as $n\rightarrow\infty$.

\subsection{Numerical results}
\label{sec:numerics}
To explore the nonlinear instability and the formation of corners of the boundary, we solve the system (\ref{eq:system}) numerically.  Our method is based on a complex variable formulation, which is an extension of well-known exact solution methods~\cite{Polubarinova-Kochina1945, Galin1945, Howison1992}.  We have detailed this numerical method elsewhere~\cite{Dallaston2013_NON, Dallaston2013_CTAC}.

Let $\boldsymbol x \in \mathbb R^2$ be represented by the complex spatial variable $z \in \mathbb C$.  To handle the free boundary we define $z=z(\zeta,t)$ to be the time-dependent mapping function from the unit disc $|\zeta|<1$ to the fluid region $\Omega(t)$.  If $\zeta=0$ maps to the far field $z=\infty$, then $z$ will have the series expansion
\begin{equation}
\label{eq:zseries}
z(\zeta,t) = \frac{a_{-1}(t)}{\zeta} + \sum_{n=0}^\infty a_n(t)\zeta^n.
\end{equation}
Laplace's equation (\ref{eq:laplace}) and the far-field condition (\ref{eq:farfield}) are satisfied by defining $\phi$ to be the real part of a complex potential $w(\zeta,t)$, analytic in the punctured disc $0<|\zeta|<1$, with a logarithmic singularity at the origin.  The dynamic condition~(\ref{eq:d}) implies $w = -\log\zeta + \mathcal V(\zeta,t)$, where $\mathcal V$ is the function analytic in the unit disc whose real part coincides with the normal velocity $v_n$:
\begin{equation}
\Re\{\mathcal V\} = v_n = \frac{\Re\{z_t\overline{\zeta z_\zeta}\}}{|\zeta z_\zeta|}, \qquad |\zeta|=1.
\end{equation}
From the kinematic condition~(\ref{eq:k}) we obtain the \emph{Polubarinova--Galin} (PG) equation
\begin{equation}
\label{eq:PG}
\Re\{z_t\overline{\zeta z_\zeta}\} = \Re\{\zeta w_\zeta\} = -1 + \Re\{\zeta \mathcal V_\zeta\}, \qquad |\zeta|=1,
\end{equation}
which is an analogue of the PG equation used in constructing exact solutions for unregularised Hele--Shaw flow~\cite{Galin1945, Howison1992,Polubarinova-Kochina1945}. Unlike the unregularised case, there are no known non-trivial exact solutions to (\ref{eq:PG}).  We instead solve (\ref{eq:PG}) numerically, by truncating the series expression (\ref{eq:zseries}) for $z$, and obtaining implicit equations for the evolution of the series coefficients $a_n$ by evaluating (\ref{eq:PG}) at equally spaced points on the circle $|\zeta|=1$.  By performing this evaluation with the fast Fourier transform, the numerical scheme is very time efficient, completing in seconds on a desktop computer even for a relatively large (256--512) number of modes.  For the numerical results in this paper we assume symmetry in $x$- and $y$-axes, meaning the coefficients $a_n$ are real and only nonzero for odd $n$.  We found this assumption greatly improved the stability of the numerical scheme.

\begin{figure}
\centering
\includegraphics[scale = 0.98]{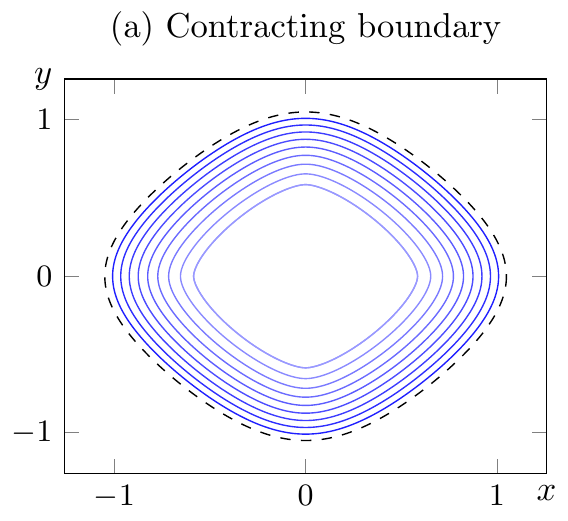}
\includegraphics[scale = 0.98]{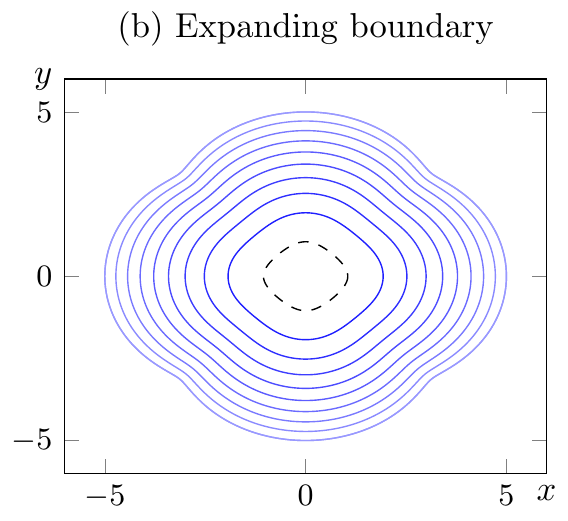}

\includegraphics[scale = 0.98]{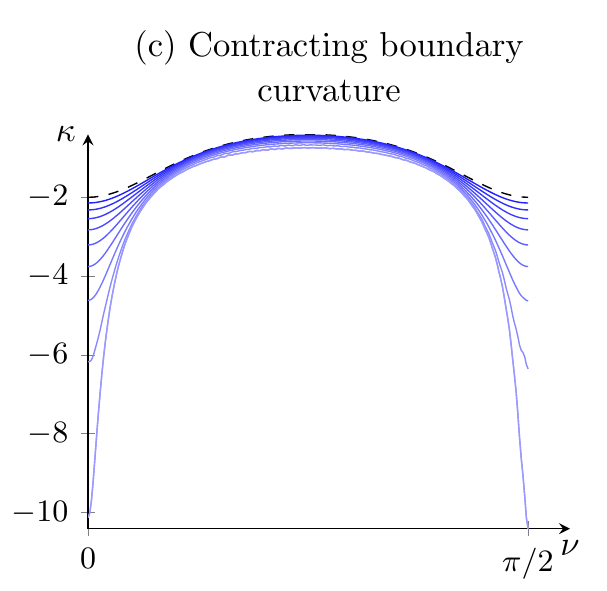}
\includegraphics[scale = 0.98]{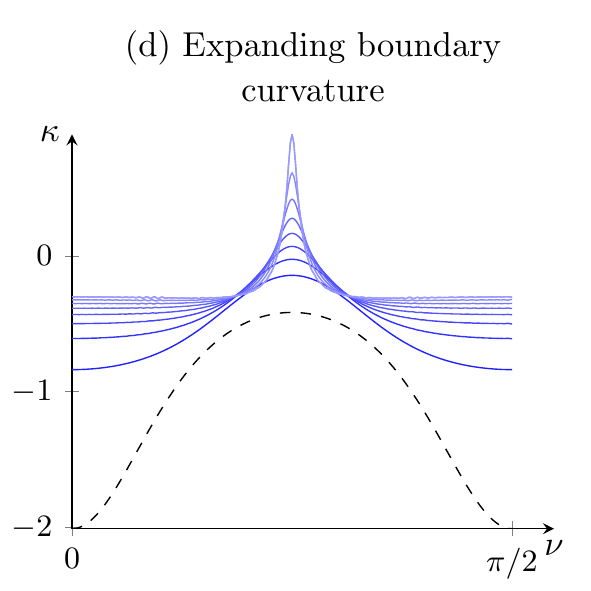}

\caption{Numerical results for contracting and expanding bubbles, showing (a,b) the boundary positions and (c,d) curvature $\kappa$ as a function of $\nu = \arg\zeta$ over the first quadrant $0 < \nu < \pi/2$, where $\zeta$ is the auxiliary complex variable.  Dashed lines represent initial conditions.  Both expanding and contracting bubbles exhibit corner formation.}
\label{fig:numerics}
\end{figure}

In Figure~\ref{fig:numerics} we present numerical solutions for an initially near-circle bubble with four-fold symmetry ($a_{-1}(0) = 1$, $a_{3}(0) = 0.05$, all other modes zero), for both contracting and expanding cases.  The boundary is unstable in both directions, as the stability analysis in Section~\ref{sec:stability} suggests.  The numerical solutions also demonstrate that the instability leads to the formation of corners, for contracting and expanding bubbles.  The curvature, also plotted in Figure~\ref{fig:numerics}, blows up at these points.  
The numerical scheme cannot be continued past the time of corner formation, as the corners correspond to singularities on the unit circle $|\zeta|=1$.

For expanding bubbles, these numerical results demonstrate the generic tendency of kinetic undercooling to cause curvature singularities in the form of corners, while preventing blow-up in the velocity of the speed, which is the generic behaviour of the unregularised problem.  This contrasts with the regularising effect of surface tension, which suppresses blow-up of both curvature and speed (see Table~\ref{tab:table}).

In the case of contracting bubbles, the numerical results demonstrate the instability predicted by the linear stability analysis in Section~\ref{sec:stability}.  The numerical results suggest this instability also leads to corner formation.  In the next section we show that, to leading order, the boundary of a small bubble evolves with constant normal velocity (up to a rescaling of time).  An exact solution to the leading order problem provides further insight into the formation of corners, and the subsequent evolution and extinction of the bubble.

\section{The small bubble limit}
\label{sec:smallbubble}
In this section we consider the behaviour of the system (\ref{eq:system}) for the case in which the (dimensionless) size of the bubble $\beta(t)$ is much less than unity; this is equivalent to a dimensional bubble size much smaller than the kinetic undercooling parameter $c$.  This regime is ultimately attained by all contracting bubbles that shrink to a point, but is also valid for small expanding bubbles (which are stable at this scale, according to the stability analysis of the previous section).

In the small bubble limit, kinetic undercooling is the dominant effect on the boundary, as opposed to the far-field source, as we demonstrate using formal asymptotics.  Let $l(t)$ be some characteristic length scale of the contracting bubble, so that $l\rightarrow 0^+$ as the bubble tends to extinction.  Let $\boldsymbol x = l \boldsymbol X$, $\phi = l^{-1}\Phi$, and define $T = -\log l$ to be the new time-like parameter (note that $T\rightarrow\infty$ as $l\rightarrow 0^+$).  The system (\ref{eq:system}) becomes
\begin{subequations}
\label{eq:rescaledsystem}
\begin{align}
&\nabla^2\Phi = 0, &\boldsymbol X \in \hat\Omega, \\
&\frac{\partial\Phi}{\partial n} = l B(X_n - V_n), & \boldsymbol X \in \hat{\partial\Omega},\\
&\Phi = B(X_n - V_n), & \qquad\boldsymbol X \in \hat{\partial\Omega}, \label{eq:rescaledkbc}\\
&\Phi \sim -l \log|\boldsymbol X| + P(T), & \qquad|\boldsymbol X|\rightarrow \infty,
\label{eq:rescaledff}
\end{align}
\end{subequations}
where spatial derivatives are now with respect to rescaled variables, $V_n$ is the rescaled normal velocity, $X_n = \boldsymbol X\cdot \boldsymbol n$, and
\begin{equation}
B = B(T) = \frac{1}{2}\frac{\mathrm d}{\mathrm dt}\left(l^2\right).
\end{equation}
The constant rate of change (\ref{eq:area}) in the area $\mathcal A$ of the bubble $\beta(t)$ implies $B = \mathcal O(1)$.  Indeed, we could define $l = \sqrt{\mathcal A}$, in which case (\ref{eq:area}) implies $B=-\pi$.  The $\mathcal O(1)$ term $P$ in (\ref{eq:rescaledff}) is included as it becomes dominant in the small bubble limit $l\rightarrow 0$ for any fixed $\boldsymbol X$. To leading order in $l$, the evolution of the boundary is dominated by kinetic undercooling on the boundary, while the fluid velocity in $\hat{\Omega}$ vanishes; if $\Phi \sim \Phi_0 + \mathcal O(l)$, then $\Phi_0$ satisfies the homogeneous Neumann problem
\[
\nabla^2\Phi_0 = 0, \qquad \left.\frac{\partial\Phi}{\partial n}\right|_{\boldsymbol X \in \hat{\partial\Omega}}=0, \qquad \Phi_0\big|_{|\boldsymbol X|\rightarrow \infty} \rightarrow P(T).
\]
The solution is trivially $\Phi_0 = P(T)$.  The kinematic condition (\ref{eq:rescaledkbc}) is therefore 
\begin{equation}
\label{eq:constvel00}
V_n = X_n + \frac{P(T)}{B(T)}.
\end{equation}
 In the original unscaled variables ($\boldsymbol x$, $t$) the leading order evolution of the interface (\ref{eq:constvel00}), in the limit it contracts to a point, is thus
\begin{equation}
\label{eq:constvel}
v_n = p(t), \qquad \boldsymbol x \in \partial\Omega.
\end{equation}
The time-dependent funtion $p(t)$ can be taken to be unity under the appropriate time transformation $t\mapsto\tilde t$ (we do not consider this time rescaling further, as it is non-trivial, but it may be found, for example, by enforcing the constant area decrease~(\ref{eq:area})).  We therefore obtain the constant velocity equation:
\begin{equation}
\label{eq:constvel0}
v_n = 1, \qquad \boldsymbol x \in \partial\Omega.
\end{equation}
Exact solutions for a boundary evolving according to (\ref{eq:constvel0}) are readily found, and the possibility of corner formation is known~\cite{Sethian1999}.  In Section~\ref{sec:exact} we construct the exact solution for an initially elliptical boundary, in which corners form, and show that it accurately approximates the numerical solution to the full Hele--Shaw problem.

\subsection{Exact solution for an ellipse}
\label{sec:exact}
In Cartesian coordinates, $v_n = 1$ is equivalent to the PDE $y_{\tilde t} = -\sqrt{1+y_x^2}$, where $(x,\pm y(x,\tilde t))$ are points on the moving boundary (assumed to be symmetric in the $x$-axis).  By differentiating in $x$ we find a first order quasilinear PDE for $y_x = y'$:
\begin{equation}
y'_{\tilde t} + \frac{y'}{\sqrt{1+y'^2}}y'_x = 0.
\end{equation}
We solve this equation using the method of characteristics.  If $x = x(x_0, \tilde t)$ is a characteristic curve originating at $x_0$, then $y'(x,\tilde t)$ is constant and equal to $y'_0(x_0)$, where $y'_0(x) = y'(x,0)$ is the initial condition. The characteristics are straight lines satisfying
\begin{equation}
\label{eq:chareq}
\frac{\partial x}{\partial \tilde t} = \frac{y'_0(x_0)}{\sqrt{1+y'_0(x_0)^2}}.
\end{equation}
Without loss of generality, we take the initial boundary to be an ellipse centred at the origin, with semimajor axis (in the $x$ direction) of unity and semiminor axis (in the $y$ direction) of $\alpha \leq 1$.  The initial condition is therefore
\begin{equation}
y(x,0) = \alpha\sqrt{1-x^2}, \qquad y'_0(x) = -\frac{\alpha x}{\sqrt{1-x^2}}.
\end{equation}
By integrating (\ref{eq:chareq}) we obtain a solution parametrised by $x_0$:
\begin{subequations}
\label{eq:charsolellipse}
\begin{align}
x &= x_0 - \frac{\alpha x_0 \tilde t}{\sqrt{1-(1-\alpha^2)x_0^2}}, \label{eq:characteristicellipse}\\
y' &= -\frac{\alpha x_0}{\sqrt{1-x_0^2}}.
\end{align}
\end{subequations}
We obtain an expression for $y$ by integrating:
\begin{equation}
\label{eq:yellipse}
y = \int y'_0\frac{\partial x}{\partial x_0} \, \mathrm dx_0 = \sqrt{1-x_0^2}\left(\alpha - \frac{\tilde t}{\sqrt{1-(1-\alpha^2)x_0^2}}\right).
\end{equation}
 (The initial condition and the requirement $y_{\tilde t} = -1$ at $x_0=0$ is enough to remove the arbitrary constant.)  The vertical coordinate $y$ has zeros at $x_0 = \pm 1$ and $x_0 = \pm x_0^*$, where
\begin{equation}
x_0^* = \sqrt{\frac{1-(\tilde t/\alpha)^2}{1-\alpha^2}}.
\end{equation}
A shock forms in the solution (\ref{eq:charsolellipse}) when $\tilde t=\alpha^2$, after which $x$ ceases to be invertible with respect to $x_0$ in the interval $[-1,1]$ (corresponding to the overlap of characteristic curves).  After this time we also have $x_0^* < 1$.  The boundary is therefore given by the parametric equations (\ref{eq:characteristicellipse}), (\ref{eq:yellipse}) for $x$ and $y$, over the interval
\begin{equation}
-\min\{1,x_0^*(\tilde t)\} \leq x_0 \leq \min\{1,x_0^*(\tilde t)\}.
\end{equation}
Corners form on the $x$-axis at $\tilde t_\text{cnr} = \alpha^2$ and at the point $x_\text{cnr} = x(1,\alpha^2) = 1-\alpha^2$.

The initial condition generalises to an ellipse with semi-axes of $b$ and $\alpha b$ by the rescaling $x \mapsto bx$, $y \mapsto by$, $\tilde t \mapsto b\tilde t$.  In this case, the time and point of corner formation are
\begin{equation}
\label{eq:cornertime}
\tilde t_\text{cnr} = b\alpha^2, \qquad x_\text{cnr} = b(1-\alpha^2).
\end{equation} 
The boundary contracts to the origin at the extinction time $\tilde t_\mathrm{ext} = \alpha b$.  The corner angle and the aspect ratio of the boundary vanish in this limit, so the boundary is asymptotically a slit in shape.  This is reminiscent of the generic \emph{eye-closing} behaviour of contracting level sets of the eikonal equation~\cite{Angenent2004} (see the discussion in Section~\ref{sec:discussion}).

\subsection{Comparison to a shrinking ellipse in Hele--Shaw flow}
To solve the Hele--Shaw problem (\ref{eq:system}) for an ellipse, we use our numerical scheme described in Section~\ref{sec:numerics} with the conformal mapping function
\begin{equation}
f(\zeta,0) = \frac{b}{2}(1+\alpha)\frac{1}{\zeta} + \frac{b}{2}(1-\alpha)\zeta
\end{equation}
as an initial condition.  This mapping transforms the unit disc to the exterior of the ellipse with semi-axes $b$ and $\alpha b$.  If $b \ll 1$, then the exact solution  (\ref{eq:characteristicellipse}), (\ref{eq:yellipse}) to the constant velocity problem (\ref{eq:constvel0}), with lengths scaled by $b$, is a good approximation to the numerical solution, up to the time of corner formation.  The only complication is the time rescaling $t \mapsto \tilde t$ introduced to obtain the constant velocity problem (\ref{eq:constvel0}) from (\ref{eq:constvel}).

\begin{figure}
\includegraphics{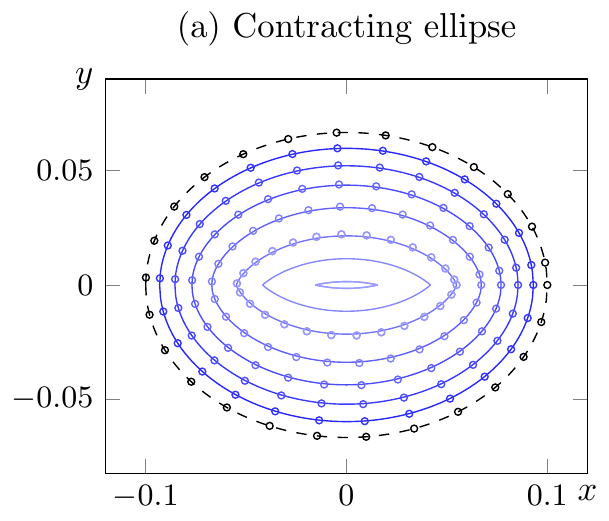}
\includegraphics{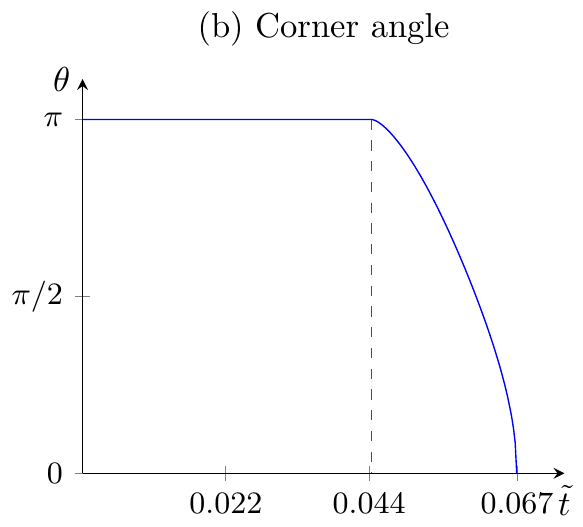}

\caption{(a) The numerical solution (circles, see Section~\ref{sec:numerics}) for a contracting bubble with kinetic undercooling, compared to the exact solution (\ref{eq:characteristicellipse}), (\ref{eq:yellipse}) (solid lines) to the constant velocity problem (\ref{eq:constvel}).  In both cases the initial condition (dashed line) is an ellipse of aspect ratio $\alpha = 2/3$ and semimajor axis $b = 1/10$.  Since the time scales are different for each solution, the times for the exact solution are chosen such that the $x$-coordinates match with the numerical solution.  The corners form at a given point (\ref{eq:cornertime}), (\ref{eq:approxcorner}), after which only the exact solution, with persistent corners, may be continued.  (b) The corner angle $\theta$ of the exact solution, starting at $\pi$ (no corner), and ending at zero.  The corner angle is a once (but not twice) differentiable function of time $\tilde t$.}
\label{fig:charsolellipse}
\end{figure}

In Figure~\ref{fig:charsolellipse} we compare the exact constant velocity solution and numerical Hele--Shaw solution for an elliptic initial condition with $b=1/10$ and $\alpha  = 2/3$.  As well as the apparent visual agreement between the two solutions, we numerically approximate the time of corner formation in the Hele--Shaw problem by using the time at which the scheme breaks.  Using $512$ terms in the expansion (\ref{eq:zseries}) we obtain
\begin{equation}
\label{eq:approxcorner}
t_\text{cnr} \approx 0.00275, \qquad x_\text{cnr} \approx 0.0548.
\end{equation}
While we cannot directly compare the time of corner formation, the point $x_\text{cnr}$ compares well with the value of $b(1-\alpha) = 0.0555\ldots$ predicted by the constant velocity solution (\ref{eq:cornertime}).  We also plot the corner angle $\theta$ of the exact solution, defined by
\begin{equation}
\theta = -2\arctan\left.\left(\frac{\partial y}{\partial x_0} \bigg/ \frac{\partial x}{\partial x_0}\right)\right|_{x_0=\min\{1,x_0^*\}}.
\label{eq:cornerangle}
\end{equation}
From this expression it can be shown that the corner angle is continuous and differentiable with respect to time $\tilde t$ at $\tilde t_\mathrm{cnr}$, although it is not twice differentiable.  

\section{Fingering and corners in a channel}
\label{sec:channel}
In this section we consider the effect of kinetic undercooling on the evolution of an initially perturbed planar front moving into the viscous fluid, using numerical solutions to the system (\ref{eq:channelsystem}) for varying kinetic undercooling coefficients $c$.  The planar front has been previously shown to be unstable~\cite{Chapman2003, Howison1992}; as with the expanding bubble, all modes of perturbation are unstable, although the inclusion of kinetic undercooling moderates their growth as mode number $n$ increases, suppressing the formation of cusp singularities.  Depending on the magnitude $c$, there are two possible long-time behaviours the boundary may tend toward:
\begin{enumerate}
\item Travelling fronts that extend across the whole channel width, which typically have corners;
\item Travelling fingers (of unbounded length) that take up some fraction $\lambda$ (that is, of nondimensional width $2\lambda$) of the channel.
\end{enumerate}
In this section we reproduce the exact solutions for the front~\cite{Chapman2003}, and outline a numerical scheme for the travelling finger with kinetic undercooling (first appearing in our conference paper~\cite{Dallaston2011}).  This scheme shows that a continuous family of finger solutions exists with widths $\lambda \in [\lambda_\mathrm{min},1)$, given a kinetic undercooling $c$; the minimum width $\lambda_\mathrm{min} \rightarrow 0$ as $c\rightarrow 0$.  These results explain previous numerical results~\cite{Romero_phd}, while suggesting the selection mechanism described by the asymptotic analysis~\cite{Chapman2003} is more subtle than the surface tension case, and is not picked up by the numerical scheme outlined here.

To close this section we consider the time-dependent problem in a channel.  Chapman and King~\cite{Chapman2003} hypothesise that travelling waves of the first kind (fronts with corners) are stable for sufficiently large kinetic undercooling compared to the channel width ($c >1$), while travelling waves of the second kind (fingers) are stable for small kinetic undercooling $c<1$.  We use our numerical scheme described in Section~\ref{sec:numerics} to support this hypothesis.

\subsection{Travelling fronts}
Exact travelling wave solutions of the Hele--Shaw problem in a channel (\ref{eq:channelsystem}) exist in the form~\cite{Chapman2003}
\begin{equation}
\phi = x-t+c, \qquad f(y,t) = t + g(y),
\label{eq:frontsoln}
\end{equation}
where $x=f(y,t)$ on the boundary $\partial\Omega$.  The ansatz~(\ref{eq:frontsoln}) identically satisfies Laplace's equation~(\ref{eq:channellaplace}), the kinematic condition~(\ref{eq:channelk}), the wall conditions~(\ref{eq:channelwalls}) and the far field condition~(\ref{eq:channelfarfield}).  The dynamic condition~(\ref{eq:channeld}) gives an equation for $g$:
\begin{equation}
g + c = \frac{c}{\sqrt{1+g'^2}}.
\label{eq:g}
\end{equation}
For $c \geq 1$, the solutions of (\ref{eq:g}) are one parameter family of circular arcs
\begin{equation}
g = \sqrt{c^2 - (y-y_0)^2} + c.  
\end{equation}
Allowing the existence of corners, a travelling front may be made of a piecewise combination of these solutions with different values of $y_0$.  In this section, we restrict ourselves to the evolution of a boundary with an initial perturbation that is symmetric about the centreline of the channel, and pointing into the fluid region at the centreline (as depicted in Figure~\ref{fig:schematic}; see also the initial conditions depicted in Figure~\ref{fig:channel}).  In Section~\ref{sec:numnumchannel} we show numerically that the perturbation spreads out to the channel walls, where the corners form.  The relevant travelling front in this case is thus
\begin{equation}
g(y) = \sqrt{c^2 - y^2} - c,
\label{eq:front}
\end{equation}
which has corners at the channel walls at $y\pm 1$.

\subsection{Numerical results for travelling fingers}
{
Here we outline a numerical solution method for computing the shape of a travelling finger in Hele--Shaw flow with kinetic undercooling.  This method is based on numerical approaches to the equivalent problem with surface tension}~\cite{Mclean1981,Vandenbroeck1983}.  One variation of the scheme {is to allow for the possibility of an artificial corner at the nose and let the width parameter $\lambda$ be arbitrary; there then exist only discrete values of the $\lambda$ for which the artificial corner angle vanishes}~\cite{Vandenbroeck1983}.  The selected finger widths $\lambda \rightarrow \half^+$ as the surface tension vanishes, which is corroborated by exponential asymptotic analysis of the same problem~\cite[e.g.]{Chapman1999}.
In contrast, our numerical results for the problem with kinetic undercooling do \emph{not} provide a clear selection of discrete finger widths, instead allowing for corner-free fingers of any width above a certain value $\lambda_\mathrm{min}$, which vanishes as the kinetic undercooling parameter $c\rightarrow 0$.  Given that asymptotic analysis~\cite{Chapman2003} predicts analytic solutions selects discrete widths, almost all the corner-free solutions we compute must be non-analytic in some other, more subtle fashion.

Taking a reference frame in which the finger is stationary, the fluid region is mapped to the upper half $\zeta$-plane by $\zeta = \exp{\pi (w-w(0))}$, where $w$ is the complex potential.  The fluid log-speed $\log q$ and velocity angle $\theta$ are harmonic conjugates, and on the real line $\zeta = \xi\in\mathbb R$, $\theta$ is only nonzero on the interval $[0,1]$.  The dynamic condition (\ref{eq:d}) becomes a nonlinear ordinary differential equation relating $q$ and $\theta$.  Allowing for a possible corner with interior angle equal to $-2\thetanose$ at the nose of the finger ($\zeta = 1$), the problem reduces to the following nonlinear integrodifferential system for $q$ and $\theta$~\cite{Chapman2003}:
\begin{subequations}
\begin{align}
&2\epsilon q \cos\theta \xi \frac{\mathrm d\theta}{\mathrm d\xi} + \cos\theta - q = 0, \label{eq:integro1} \\
&\log(1-\lambda) = \frac{1}{\pi}\int_0^1 \frac{\theta(\xi)}{\xi} \ \mathrm d \xi,  \label{eq:integro2} \\
&\log q = \log(1-\lambda) - \frac{1}{\pi}\dashint_0^1 \frac{\theta(\xi')}{\xi'-\xi} \ \mathrm d \xi', \\
&\theta(0) = 0, \ \theta(1) = \thetanose,
\end{align}
\label{eq:integro}
\end{subequations}
where $\epsilon$ is related to the kinetic undercooling parameter $c$ by
\begin{equation}
\epsilon =  \frac{c\pi}{2(1-\lambda)}.
\label{eq:defeps}
\end{equation}
For numerical efficiency, endpoint singularities in $\xi$ are removed by introducing the new independent variable $t$ defined by
\begin{equation}
\label{eq:tdef}
t = \sqrt{1-\xi^\alpha},
\end{equation}
where the exponent $\alpha$ is found by solving a transcendental equation (see our previous work~\cite{Dallaston2011} for further details).  This transformation also leads to a greater density of points near the endpoints $\xi = 0$ and $\xi = 1$, corresponding to the tail and nose of the finger, respectively.
For given values of $\epsilon$ and $\lambda$, we discretise the system (\ref{eq:integro}) by dividing $t \in [0,1]$ into $N+1$ grid points $t_n$, where $n = 0,1,\ldots,N$, and solving the $N\times N$ nonlinear system resulting from (\ref{eq:integro1}) and (\ref{eq:integro2}) for the unknown values of $\thetanose$ and $\theta_n$ at interior grid points ($n = 1,\ldots, N-1$). 

 To support the classical approach of a finite difference discretisation on evenly spaced gridpoints, we also implemented a spectral collocation method based on Chebyshev polynomials, where $t_n$ are chosen to be Chebyshev nodes (this leads to greater concentration of nodes near the end points) and differentiation and integration are performed in the space of coefficients, using well known algorithms~\cite[e.g.]{Fox1968}.  More details of this procedure exist in the appendix of the PhD thesis~\cite{Dallaston_thesis}.  The spectral collocation method provided the same results and performed similarly to the finite difference discretisation; while overall fewer nodes were needed in the spectral method (20--30 rather than 50--100), the run time was similar (on the order of seconds per value of $\lambda$ and $\epsilon$), and we did not attain spectral convergence, likely due to remaining weak singularities at the endpoints.

\begin{figure}
\includegraphics{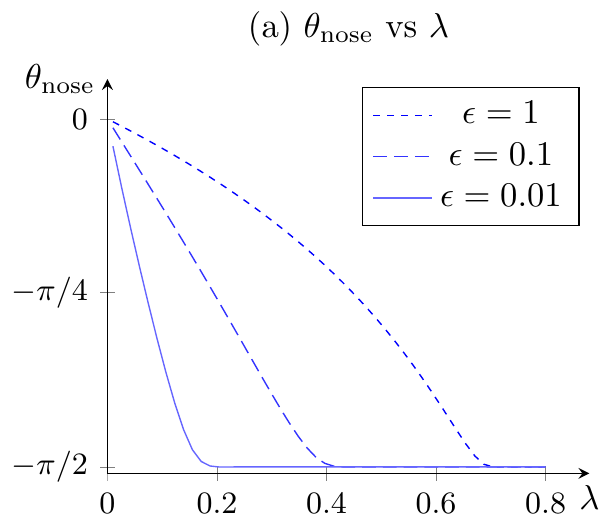}
\includegraphics{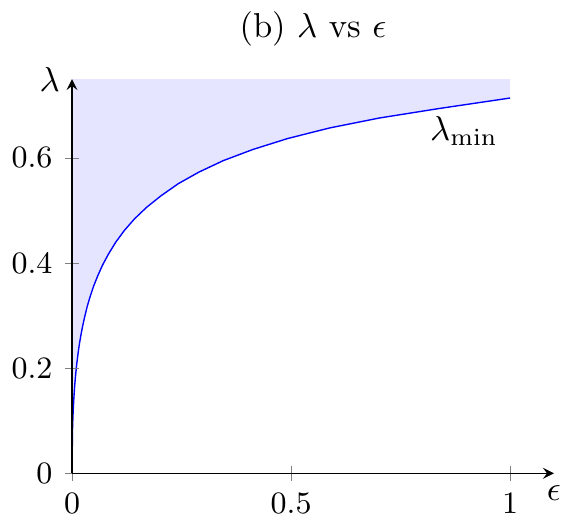}

\includegraphics{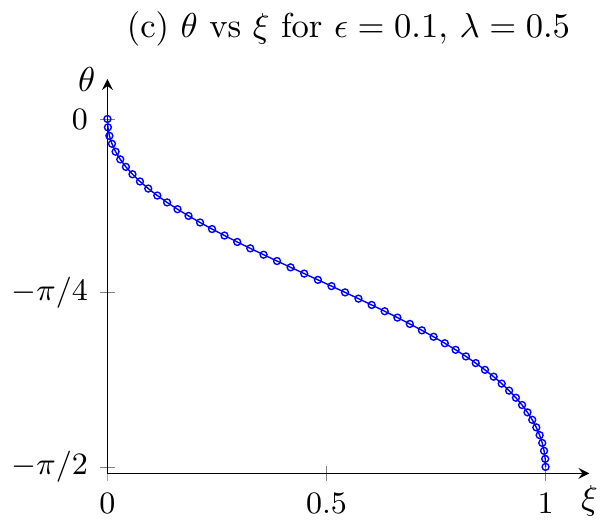}
\includegraphics{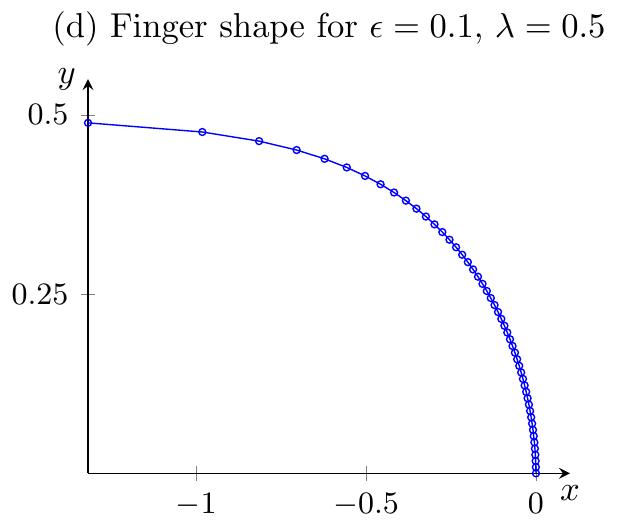}

\caption{(a) Nose angle $\thetanose$ as a function of finger width $\lambda$, for kinetic undercooling coefficient $\epsilon = 0.01$, $0.1$ and $1$.  For $\lambda$ greater than a minimum $\lambda_\mathrm{min}$, the nose angle $\thetanose = -\pi/2$, corresponding to a corner-free finger.  (b) The dependence of $\lambda_\mathrm{min}$ on $\epsilon$, with the shaded region corresponding to the existence of fingers without corners at the nose. (c) An example solution to the integrodiffierential system (\ref{eq:integro}) for $\epsilon = 0.1$, $\lambda = 0.5$, and $N=50$ nodes, showing the nodes, evenly spaced in the transformed variable $t$ (\ref{eq:tdef}), are concentrated at each end point.  (d) The shape of the example travelling finger solution in physical coordinates, also showing the distribution of nodes.}
\label{fig:vandenbroeck}
\end{figure}

A nose angle $\thetanose = -\pi/2$ corresponds to a boundary that is differentiable at the nose, with no corner.  We solve the problem (\ref{eq:integro}) for a range of $\lambda$ values, to find the dependence of $\thetanose$ on $\lambda$.  In Figure~\ref{fig:vandenbroeck}a we plot the resulting curves of nose angle $\thetanose$ against width $\lambda$, for kinetic undercooling coefficient $\epsilon = 0.01$,  $0.1$ and $1$.  For small $\epsilon$, the nose angle $\thetanose$ is close to zero, representing a sharp corner; however, as $\epsilon$ increases, it reaches the value of $-\pi/2$ at a minimum $\lambda = \lambda_\mathrm{min}(\epsilon)$, at which it remains.  Due to numerical error, very small oscillations (of the order of numerical error) in $\thetanose$ exist near $\lambda = \lambda_\mathrm{min}$ around the value of $-\pi/2$, which explains the discrete solution branch with $\lambda\rightarrow 0$ we previously observed~\cite{Dallaston2011}, roughly corresponding to $\lambda_\mathrm{min}$.  Thus there is a continuous family of corner-free finger solutions for any given $\epsilon$, in contrast to the surface tension case, where $\thetanose$ is equal to $-\pi/2$ only at discrete points~\cite{Vandenbroeck1983}. In Figure~\ref{fig:vandenbroeck}b we plot $\lambda_\mathrm{min}$ against $\epsilon$, showing that $\lambda_\mathrm{min}\rightarrow 0$ as $\epsilon\rightarrow 0$.  In Figure~\ref{fig:vandenbroeck}c and d, we show an example solution for $\epsilon=0.1$ and $\lambda = 0.5$, and the distribution of nodes that are evenly spaced in the unit interval in $t$, thus demonstrating the density of nodes near the nose of the finger.

At first glance these results seem to be at odds with the asymptotic analysis of Chapman and King~\cite{Chapman2003}, which predicts that only discrete branches of analytic fingers exist, with a countable number of allowed widths $\lambda$ for a given $\epsilon$; furthermore, each branch has $\lambda\rightarrow \half^+$ as $\epsilon\rightarrow 0^+$ (as with surface tension).  Our results reconcile with these predictions given that a boundary may have no corner at the nose but still not be analytic there; we have found the family of solutions which are corner-free (the shaded region in Figure~\ref{fig:vandenbroeck}b), but the numerical method is unable to resolve non-analytcities of a more subtle nature (for instance, nonexistence of the arclength derivative of curvature, or higher derivatives).  The complicated nature of the selection problem near the nose in the asymptotic analysis~\cite{Chapman2003}, compared to the more straightforward problem with surface tension, also supports this view.  Our numerical results also conform to those of Romero~\cite{Romero_phd}, who found no discrete selection effect from kinetic undercooling; the numerical scheme devised therein is also unlikely to discriminate between analytic solutions and those with such subtle non-analyticities.

\subsection{Numerical results for an evolving front}
\label{sec:numnumchannel}
We now adapt our numerical scheme outlined in Section~\ref{sec:numerics} to the channel geometry.  The approach is very similar, except the mapping function $z$ now takes the form
\begin{equation}
z(\zeta,t) = -\frac{1}{\pi}\log\zeta + \sum_{n=0}^\infty a_n(t) \zeta^n,
\label{eq:zserieschannel}
\end{equation}
while the kinetic undercooling coefficient $c$ is introduced to the Polubarinova--Galin equation (\ref{eq:PG}).  On the assumption the finger is symmetric around the centreline $y=0$, the coefficients $a_n$ are real.  While this solution method is not well-suited to approximating highly elongated boundaries such as those with long fingers, we found some improvement for moderately deformed boundaries could be made by composing $f(\zeta,t)$ with the M\"obius transformation (which preserves the unit disc)
\begin{equation}
\zeta = \frac{\chi + r}{1+\chi r}, \qquad 0 \leq r< 1,
\end{equation}
and solving (\ref{eq:PG}) at equally spaced point on the unit circle in the $\chi$-plane.  By altering the value of $r$, we have some control over the node spacing on the boundary, while the fast Fourier transform is still applicable.

\begin{figure}
\centering
\includegraphics{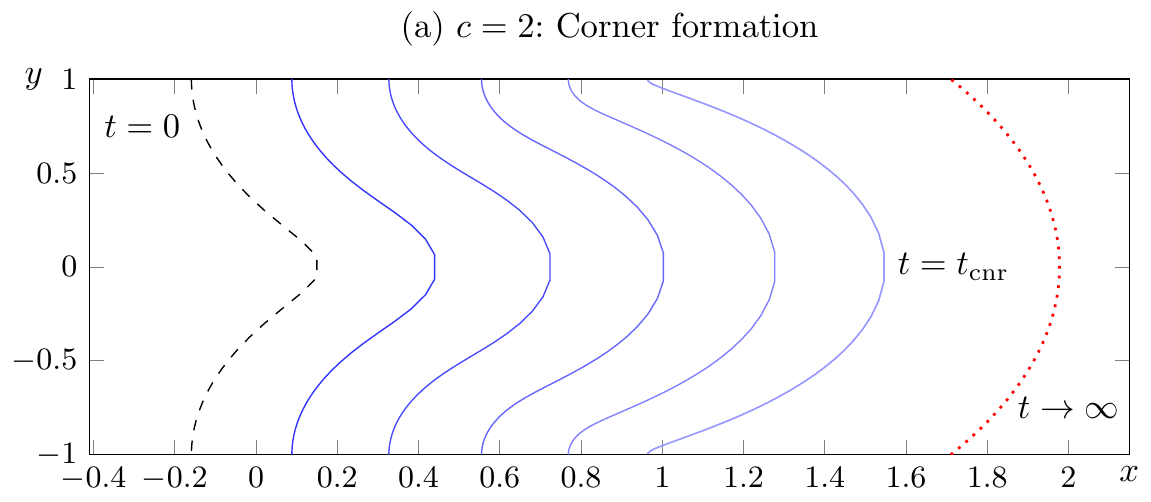}

\includegraphics{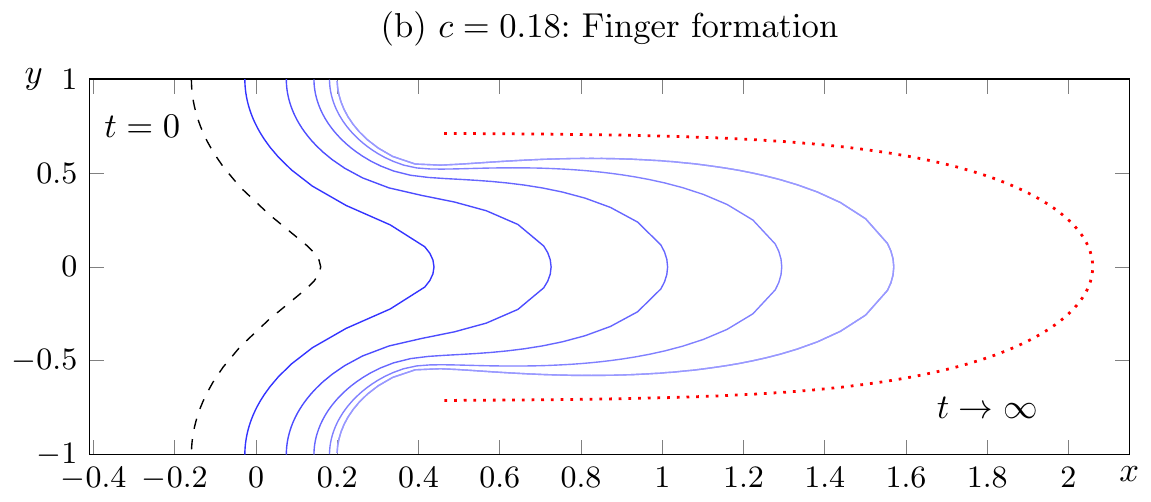}

\caption{Evolution of a perturbed boundary ($a_1 = 0.5$ in (\protect{\ref{eq:zserieschannel}})) in a Hele--Shaw channel, with kinetic undercooling regularisation.  (a) For large kinetic undercooling ($c = 2 > 1$), the initial perturbation (dashed line) grows and spreads out to the channel walls, forming corners at a finite time $t_\text{cnr}$.  The travelling wave for $c=2$, given by (\ref{eq:front}), is shown as a dotted line.  (b) For small kinetic undercooling ($c = 0.18 < 1$), the perturbation does not spread to the cell walls, instead tending to a finger of width $2\lambda$, less than the channel width.  In this case the travelling finger solution (dotted line) comes from the numerical solution of the integrodifferential system (\ref{eq:integro}) with $\epsilon=1$, and $\lambda = \lambda_\mathrm{min} = 0.71$.}
\label{fig:channel}
\end{figure}

In Figure~\ref{fig:channel} we plot the numerical solution of an initially perturbed boundary ($a_1 >0$, all other modes zero), for two values of the kinetic undercooling coefficient $c$: for $c = 2 > 1$, we expect the travelling front (\ref{eq:front}) to be stable, while for $c = 0.18 < 1$, we expect the boundary to tend to an infinitely long finger given by the numerical solution of (\ref{eq:integro}).  Here the value of $c=0.18$ is chosen to correspond (using (\ref{eq:defeps})) to $\epsilon = 1$, and $\lambda = \lambda_\mathrm{min} = 0.71$ is the computed minimum width for which a finger exists that has no corner at the nose (note, however, this finger may not be analytic and it is not clear which travelling finger solution is stable in the time-dependent problem; see the discussion in Section~\ref{sec:discussion}).

In each case, 256 modes are used, and the transformation parameter $r=0.8$ concentrates nodes at the channel walls.  For $c=2$, the initial perturbation grows and spreads out toward the walls, with corners forming at the walls at a finite time $t_\mathrm{cnr}$.  The numerical scheme cannot continue past this time; it is likely that the travelling front (\ref{eq:front}) (also plotted) is stable and the boundary evolves towards it, although a sophisticated numerical scheme that can handle persistent corners is necessary to test this conjecture.

For $c = 0.18$, the initial perturbation also grows and initially spreads outward, but does not cause corners to form on the walls; instead, a finger forms in the centre of the channel.  The numerical scheme breaks as the boundary points (equally spaced in the $\chi$-plane) become sparse near the finger nose and channel walls in the physical plane.  Again, more sophisticated numerical methods are needed to track the finger further, as we discuss in the following section.

\section{Discussion}
\label{sec:discussion}
We have examined the effect of kinetic undercooling on growing and contracting inviscid bubbles in Hele--Shaw flow. Stability analysis, numerical solutions and explicit solutions for the leading order approximation in the small bubble limit demonstrate the difference between kinetic undercooling, surface tension and the unregularised problem, particularly with regard to the formation of corners at finite time for both expanding and contracting bubbles in the kinetic undercooling case.

Additionally, we have numerically examined finger and corner formation of Hele--Shaw flow in a channel, demonstrating corners form for sufficiently high kinetic undercooling ($c>1$), and fingers form for lower kinetic undercooling ($c<1$).  We have examined the effect of kinetic undercooling in the problem of selecting discrete analytic solutions to the problem of a travelling finger shape. Kinetic undercooling allows a continuous family of corner-free fingers with widths above a minimum width $\lambda_\mathrm{min}(\epsilon)$, for a given kinetic undercooling strength $\epsilon$.  It is likely, given asymptotic analysis~\cite{Chapman2003}, that this continuous family is not analytic apart from discretely selected widths $\lambda$, but this non-analyticity is more subtle than the presence of a corner at the nose.  The selection cannot be resolved by numerical methods employed to date.

One area we believe deserves more attention is the generic extinction behaviour of closed curves contracting with constant normal velocity (\ref{eq:constvel0}).  The general rule of curve evolution by constant velocity is well-known, being equivalent to Huygen's principle from optics, and also used in image analysis and etching/deposition~\cite[e.g.]{Sethian1999}, but we are not aware of specific studies into the extinction behaviour of closed curves contracting to a point according to such a rule.
The exact solution presented in Section~\ref{sec:smallbubble} demonstrates two characteristics of such a curve: the formation of corners, and the slit-type (vanishing aspect ratio) extinction shape.  It is reasonable to conjecture that these two properties are generic for curves contracting according to (\ref{eq:constvel0}), and therefore for bubbles contracting in Hele--Shaw flow with kinetic undercooling.  Another issue is the possibility of self-intersection of the boundary (or \emph{pinch-off} of bubbles in the Hele--Shaw context), and which initial conditions lead to such self-intersection.  Since the curve must be non-convex to intersect with itself, we conjecture that a curve avoids such self-intersection if and only if it is initially convex.  The generic extinction shape of contracting plane curves is also a topic in focusing/hole-closing problems in curvature-driven flow~\cite{Gage1986, Grayson1987}, and the porous medium and eikonal equations~\cite{Angenent1995,Angenent2004}.  In particular, the generic \emph{closing eye} solution for the eikonal equation~\cite{Angenent2004} also exhibits the finite-time corner formation and slit-type extinction we see for (\ref{eq:constvel0}).  Our problem is likely to be related.

There is also clear scope for the application of more sophisticated numerical schemes to the full Hele--Shaw problem with kinetic undercooling.  Our approach, based on conformal mapping functions, is limited in that solutions cannot be continued past the time of corner formation, and is not particularly effective at capturing highly elongated boundaries (such as those with long fingers).  Many numerical techniques have been applied to growing Hele--Shaw bubbles and channels with surface tension, effectively capturing fingering and tip-splitting phenomena~\cite{Degregoria1986, Bensimon1986, Tryggvason1983, Ceniceros1998, Hou1997,Hou1994}.  A similar method applied to the problem with kinetic undercooling would be valuable to demonstrate the evolution of a finger in a channel and determine if it evolves from a variety of initial conditions to a steadily translating finger of unique width; this finger would likely correspond to the lowest (stable) of the discrete branches selected by their analyticity~\cite{Chapman2003}.  For large kinetic undercooling, a numerical approach that can capture corner formation and subsequent evolution is a greater challenge, although level set methods~\cite[e.g.]{Sethian1999} have demonstrated this ability.

A numerical method that allows for corners could also be used to test the stability of corners in the case of expanding bubbles.  In a channel, the boundary develops either corners or fingers depending on the ratio of kinetic undercooling to channel width, which is fixed.  For an expanding bubble, the only length scale is the bubble size, which is increasing.  One possibility in the bubble geometry is that corners are initially stable (as seen in the numerical solution in Figure~\ref{fig:numerics}) when the bubble size is of the same order as kinetic undercooling, but when the bubble is much larger, the corners become unstable and fingering solutions emerge instead.

Finally, the discrete selection of analytic fingers by kinetic undercooling, predicted by asymptotic analysis~\cite{Chapman2003}, has not yet been demonstrated numerically.  From our results, and the nature of the asymptotic problem, it is likely that the difference between an analytic solution and non-analytic (but corner-free) solution is very subtle, and a numerical method that can detect such a difference will be difficult to devise.  An alternative approach may be to add finite surface tension (eliminating the possibility of corner-free but non-analytic fingers~\cite{Tanveer2003, Xie2003}) and observe what solution is selected in the limit that surface tension vanishes.  We leave this problem for future study.

\end{document}